\documentclass[a4paper,12pt]{article}
\usepackage{epsf,amsmath}
\usepackage{graphicx}
\usepackage{amssymb}
\usepackage{array,tabularx}
\usepackage{indentfirst}
\usepackage[utf8]{inputenc}
\begin{document}

\title{Cosmological dynamics in six-order gravity}

\author{M.Skugoreva$^1$, A.Toporensky$^1$ and P.Tretyakov$^2$}
\date{}
\maketitle
$^1$ {\it Sternberg Astronomical Institute, Moscow 119992, Russia.}

$^2$ {\it Joined Institute for Nuclear Research, Dubna, Moscow Region, Russia.}

\begin{abstract}
We consider cosmological dynamics in
generalized modified
gravity theory with the $R\Box R$ term added to the
action of the form
$R+R^N$. Influence of $R \Box R$ term to the known solutions
of modified gravity is described.
We show that in particular case of $N=3$ these two non-Einstein
terms are equally important on power-law solutions.
These solutions and their stability have been studied using
dynamical system approach. Some results for the case of
$N \ne 3$ (including stability
of de Sitter solution in the theory under investigation)
have been found using other methods.

\end{abstract}

\section{Introduction}
Theories of modified gravity become very popular in the beginning of our century
and the main reason is attempts to understand accelerated expansion of the Universe
which is proved observationally \cite{Perlmutter, Riess}. Modifying left-hand side of
Einstein equations of gravity it is in principal possible to obtain a late-time acceleration
without invoking any new type of matter. It is also known that such theories can
explain inflation (i.e. accelerated expansion at early stage of Universe evolution)
using only gravitational sector of the theory \cite{Starobinsky,Whitt}, though inflation
driven by the scalar field without any corrections to Einstein gravity is much popular
now (nevertheless, inflation from modified gravity is not currently ruled out by
observations \cite{dark}).

Standards way to modify Einstein gravity is to replace the curvature scalar $R$
in the gravity Lagrangian by some scalar function $f(R)$. In principal, it is possible
to consider Lagrangians which depends on other curvature invariants, like
the Kretchmann invariant $R_{ijkl}R^{ijkl}$ or Ricci square $R_{ik}R^{ik}$ etc.
However, cosmological evolution in such theories often lead to anisotropic
expansion \cite{Hervik, we, Barrow} and, therefore, can not describe late stages of our
Universe. This is the main reason why nowadays the class of $f(R)$ theories is the most popular
class of modified gravity, and we restrict ourselves by function of scalar curvature invariant
in the present paper (it should be noted, however, that theories with scalar function $f(G)$
of Gauss-Bonnet invariant $G=R_{ijkl}R^{ijkl}-4R_{ij}R^{ij}+R^2$ are promising due to
special properties of the invariant $G$ \cite{Odintsov1,Odintsov2}).

Apart from considering scalar function of $R$ it is possible to introduce terms depending
on derivatives of the scalar curvature. These theories are known from the end 80th,
in particular, theories with terms which are polynomial in the combination $R\Box R$
have been intensively investigated \cite{Schmidt1, Schmidt2, Amendola, Wands, KS, Odintsov3}.

The goal of the present paper is to consider a theory which contains {\it both} $f(R)$ and
$R \Box R$ modifications of the Einstein gravity. We concentrated our attention on the liner with respect to $R\Box R$ term case, because as it was shown only liner case is the ghost free one \cite{Ovrut}. We apply theory of dynamical system to
study asymptotic regimes and their stability. This method is used actively for finding asymptotic
solutions in $f(R)$ gravity and studying their stability \cite{Polarski, Dunsby1, Dunsby2}, we extend it
for theories with $R \Box R$ term.
We show that this approach works well in the situation
when corrections from these two ways of modifying gravity are equally important.
One such case is identified ($f(R)=R+\beta R^3$)
and investigated systematically. After, we discuss some problems
of the used method  in a general situation.   Finally, some modifications of known
cosmological behavior in $f(R)$ theories due to presence of $R \Box R$ term are described
for a broader class of  functions $f(R)$.

\section{Main equations}
We consider Lagrangian in the form
\begin{equation}
L=\sqrt{-\mathrm{g}}f(R,R\Box R\equiv A)+L_M.
\end{equation}
Equations of motion for such kind of theories have the form
\begin{equation}
\begin{array}{l}
-\frac{1}{2}f\mathrm{g}_{ik}+f_RR_{ik}-\nabla_i\nabla_k f_R+
\mathrm{g}_{ik}\Box f_R+\\
\\
+f_A\Box RR_{ik}+ \Box (f_AR)R_{ik}-
 \nabla_i\nabla_k(f_A\Box R+\Box(f_AR))+\Box(f_A\Box R
+\Box(f_AR))\mathrm{g}_{ik}+\\
\\
+\frac{1}{2}\nabla_l(f_AR)\nabla^lR\mathrm{g}_{ik}-\nabla_i(f_AR)\nabla_jR
+\frac{1}{2}f_AR\Box R\mathrm{g}_{ik}=8\pi GT_{ik},
\end{array}
\label{a1}
\end{equation}
and its trace
\begin{equation}
\begin{array}{l}
-2f+f_RR+3\Box f_R+Rf_A\Box R+R\Box(f_AR)+\\
\\+3\Box(f_A\Box R
+\Box(f_AR))+\nabla_l(f_AR)\nabla^lR+2f_AR\Box R=8\pi GT,
\end{array}
\label{a2}
\end{equation}
where $f_{R}=\frac{\partial f}{\partial R}$,
    $f_{A}=\frac{\partial f}{\partial A}$,
     $\nabla_iR=\frac{\partial R}{\partial x^i}$.
Since we interested in the flat Friedmann-Robertson-Walker metrics
\begin{equation}
ds^2=dt^2-a^2(t)dl^2,
\end{equation}
the following simple relations can be written down
\begin{equation}
T=\rho-3p,\,T_{00}=\rho,\,R=-6(2H^2+\dot H),\,R_{00}=\frac{1}{2}R+3H^2,\,\Box R=3H\dot R+\ddot R.
\label{a3}
\end{equation}
Denoting terms depending on
$R$ and $R\Box R$ as
$f_1$ and $f_2$ respectively, we get
\begin{equation}
f(R,\,R\Box R\equiv A)=f_1(R)+f_2(R\Box R)
\end{equation}
As it was already noted in the present paper we study the liner case of
$f_2(R\Box R)=\alpha R\Box R$, the function  $f_1(R)$ will be specified later.
It is useful to introduce the following notations: $\partial f_1/\partial R = F$ and $\alpha \Box R=\Phi$.
 As  $F$
and $\Phi$ are scalars, in analog with eq. (\ref{a3}) we get:
\begin{equation}
\Box F=3H\dot F+\ddot F,\,\,\,\,\,\,\Box \Phi=3H\dot \Phi+\ddot \Phi.
\label{a4}
\end{equation}

Taking into account this relations (00) component of eq. (\ref{a1}) may by written as
\begin{equation}
-\frac{1}{2}f_1+\frac{1}{2}RF+3H^2F+3H\dot{F}
+2(\frac{1}{2}R+3H^2)\Phi+6H\dot{\Phi}-\frac{\alpha}{2}{\dot{R}}^2=K\rho,
\label{a5}
\end{equation}
and its trace (\ref{a2}) take the form
\begin{equation}
-2f_1+FR+9H\dot{F}+ 3\ddot{F}+2\Phi R+18H\dot\Phi+6\ddot\Phi+\alpha{\dot{R}}^2=K\rho(1-3w),
\label{a6}
\end{equation}

where we denote $K=8\pi G$ and $p=w\rho$.

For further investigations we introduce the following dimensionless variables
\begin{equation}
\begin{array}{l}
x=-\frac{f_1}{6FH^2},\\
\\
y=\frac{R}{6H^2},\\
\\
z=\frac{\dot{F}}{FH}\\
\\
m=\frac{2\Phi}{F},\\
\\
n=\frac{2\dot{\Phi}}{FH},\\
\\
r=-\frac{\alpha{\dot{R}}^2}{6FH^2},\\
\\
\Omega=\frac{K\rho}{3FH^2}.
\end{array}
\label{a7}
\end{equation}
 Dividing eq. (\ref{a5}) by $3FH^2$, we rewrite it in our new variables (\ref{a7}):
\begin{equation}
\begin{array}{l}
x+y+z+1+m(y+1)+n+r=\Omega .
\end{array}
\label{a8}
\end{equation}
Similarly,
eq. (\ref{a6}) can be rewritten as
\begin{equation}
\begin{array}{l}
4x+2y+3z+\frac{\ddot F}{FH^2}+\frac{2\ddot\Phi}{FH^2}+2my+3n-2r=\Omega (1-3w) .
\end{array}
\label{a8.1}
\end{equation}

Taking derivative of the introduced variables with respect to time we get the system of equations

\begin{equation}
\begin{array}{l}
\frac{\dot{x}}{H}=-\frac{\dot{R}}{6H^3}-xz-2x\frac{\dot{H}}{H^2},\\
\\
\frac{\dot{y}}{H}=\frac{\dot{R}}{6H^3}-2y\frac{\dot{H}}{H^2},\\
\\
\frac{\dot{z}}{H}=\frac{\ddot{F}}{FH^2}-z^2-z\frac{\dot{H}}{H^2},\\
\\
\frac{\dot{m}}{H}=n-mz,\\
\\
\frac{\dot{n}}{H}=\frac{2\ddot{\Phi}}{FH^2}-nz-n\frac{\dot{H}}{H^2},\\
\\
\frac{\dot{r}}{H}=-\frac{\alpha \dot R\ddot R}{3FH^3}-rz-2r\frac{\dot{H}}{H^2},\\
\\
\frac{\dot{\Omega}}{H}=\frac{K\dot{\rho}}{3FH^3}-\Omega z-2\Omega\frac{\dot{H}}{H^2}.
\end{array}
\label{a9}
\end{equation}
To complete transformation we need to express all derivatives in (\ref{a9}) such as $\dot H$, $\dot R$ and other through new variables (\ref{a7}). First of all by using the continuity equation
\begin{equation}
\dot \rho+3H(\rho+p)=0
\label{a10}
\end{equation}
we find $\frac{K\dot{\rho}}{3FH^3}=-3\Omega(1+w)$.
Moreover since $R=-6(2H^2+\dot{H})$, it easy to find that
$\frac{\dot{H}}{H^2}=-2-y$.
Let us now introduce dimensionless parameters, characterizing the function
$f_1(R)$:
\begin{equation}
\begin{array}{l}
a=\frac{Rf_{1R}}{f_1}=\frac{-y}{x},\\
\\
b=\frac{Rf_{1RR}}{f_{1R}}=\frac{R\dot{F}}{F\dot{R}},\\
\\
c=\frac{Rf_{1RRR}}{f_{1RR}}.
\end{array}
\label{a11}
\end{equation}

 Using these parameters we easily find
$$
\frac{\dot{R}}{6H^3}=\frac{\dot{R}}{6H^3}
\frac{F}{R\dot F}\frac{F\dot F}{F}=\frac{yz}{b}.
$$

 Using (\ref{a3}) and (\ref{a4}) it is possible to calculate
$$
\frac{\ddot{F}}{FH^2}=\frac{z^2c}{b}-\frac{z^2ym}{2br}-3z.
$$

 Using expression (\ref{a8.1}) we find
$$
\frac{2\ddot{\Phi}}{FH^2}=-\frac{z^2c}{b}+\frac{z^2ym}{2br}-4x-2y-2ym-3n+2r+\Omega(1-3w).
$$

And finally, using $\ddot{R}=\frac{\Phi}{\alpha}-3H\dot{R}$ we find
$$
-\frac{\alpha \dot R\ddot R}{3FH^3}=
-\frac{yzm}{b}-6r.
$$
Note now that  using eq. (\ref{a8}) it is possible to exclude $\Omega$ from resulting system. And summing all above expressions and going to the conformal time, we find resulting system:
\begin{equation}
\begin{array}{l}
x'=-\frac{yz}{b}+x(2y+4-z),\\
\\
y'=\frac{yz}{b}+2y(2+y),\\
\\
z'=\frac{z^2c}{b}-\frac{z^2ym}{2br}-z^2+z(y-1),\\
\\
m'=n-mz,\\
\\
n'=-\frac{z^2c}{b}+\frac{z^2ym}{2br}-4x-2y-2ym+\\
\\+2r+(x+y+z+1+m(y+1)+n+r)(1-3w)+n(y-z-1),\\
\\
r'=-\frac{yzm}{b}+r(2y-2-z),
\end{array}
\label{a12}
\end{equation}
where we denote $'=\frac{d}{d(\ln a)}$.
For some functions $f_1$ there are additional relations between $x$ and $y$, a well as
between $z$ and $r$. In the next section we consider an example for which such relations exist.

\section{The case of $f_1=R+\beta R^3$.}
In this special case some simplification can be done. First of all we note that the parameter $c=1$.
Moreover, the parameter $b$ is a combination of  variables $x$ and $y$:

\begin{equation}
b=3\frac{x+y}{y}.
\label{b1}
\end{equation}
It is possible also to find relation between $r$ and $z$ in this special case:
\begin{equation}
r=-\frac{\alpha z^2y}{108\beta(x+y)}.
\end{equation}
So we can substitute this relation into
(\ref{a12}) and exclude $r$ from the system:
\begin{equation}
\begin{array}{l}
x'=-\frac{y^2z}{3(x+y)}+x(2y+4-z),\\
\\
y'=\frac{y^2z}{3(x+y)}+2y(2+y),\\
\\
z'=\frac{z^2y}{3(x+y)}+18\frac{\beta ym}{\alpha}-z^2+z(y-1),\\
\\
m'=n-mz,\\
\\
n'=-\frac{z^2y}{3(x+y)}-18\frac{\beta ym}{\alpha}
-4x-2y-2ym-\frac{\alpha z^2y}{54\beta(x+y)}+\\
\\+(x+y+z+1+m(y+1)+n-\frac{\alpha z^2y}{108\beta(x+y)})(1-3w)+n(y-z-1).
\end{array}
\label{b2}
\end{equation}
\\
As we have two coupling constants, $\alpha$ and $\beta$ our strategy is to find
stationary points and corresponding eigenvalues separately for $\alpha=1$ and
$\alpha=-1$ assuming $\beta=1$. Having this information we then consider the
general case.

We start with $\alpha=1$.
Solving the system (\ref{b2}) with vanishing left-hand sides, we find the following stationary points.
\\1. $(x,y,z,m,n,\Omega)$:
\begin{equation}
(1,  -2,  0,  0,  0,  0)
\label{b3}
\end{equation}
Corresponding eigenvalues (which have been found numerically) are
\begin{equation}
(-3-3w,-7.9040, 4.9040, -3.3677, 0.3677)
\label{b4}
\end{equation}
This is de Sitter point, and we can see that it is a saddle.

2.
\begin{equation}
\!\left ( \!\frac{1}{2}\!-\!\frac{1}{6}w,   \frac{1}{2}w\!-\!\frac{3}{2}, \! -2\!-\!2w,  \frac{(w^2\!-\!1)}{3(w-3)}, \frac{2(w^2\!-\!1)(1\!+\!w)}{3(3-w)},  \frac{(5w^3\!+\!22w^2\!-\!34w\!-\!63)}{9(3-w)} \right )
\label{b5}
\end{equation}
 At least one of the eigenvalues  is not negative
\begin{equation}
2(1+w)\geqslant 0.
\end{equation}
Other eigenvalues are
\begin{equation}
\begin{array}{l}
\\ 5/4w-1/4 \pm
\\1/4{(-7w^2-50w+313+8{(-14w^4-2w^3+313w^2-624w+387)}^{1/2})}^{1/2},\\
\\ 5/4w-1/4 \pm
\\1/4{(-7w^2-50w+313-8{(-14w^4-2w^3+313w^2-624w+387)}^{1/2})}^{1/2}.
\end{array}
\end{equation}
This is a matter dominated point in a high-curvature regime.
This point is also a saddle. Scale factor $a$ has the following dependence on time:
\begin{equation}
a(t)=a_0{|t-t_0|}^{\frac{2}{w+1}}.
\end{equation}
3.
\begin{equation}
(0.1623, -0.4870, -6.0520, -1.0625, 6.4305, 0)
\label{b6}
\end{equation}
Eigenvalues are (the last eigenvalue depends on equation of state parameter $w$,
so we present values found for several particular $w$)
\begin{equation}
( 6.4850,-1.9200, 4.5650, 6.0520, \left\{
\begin{array}{l}
9.0780, w=-1\\
7.5780, w=-0.5\\
6.0780, w=0\\
4.5780, w=0.5\\
3.0780,w=1
\end{array}
\right .)
\end{equation}

We can see that this point is a saddle. The scale factor is
\begin{equation}
a(t)=a_0{|t-t_0|}^{\frac{1}{1.5130}}.
\end{equation}
4.
\begin{equation}
(1.3732, -4.1195, 8.4778, -1.0700, -9.0710, 0)
\label{b8}
\end{equation}
Eigenvalues are
\begin{equation}
( -13.5973,-8.0759, -8.4778, -5.5213, \left\{
\begin{array}{l}
-12.7167, w=-1\\
-14.2167, w=-0.5\\
-15.7167, w=0\\
-17.2167, w=0.5\\
-18.7167,w=1
\end{array}
\right .)
\end{equation}

We can see that for all studied values of $w$ this point is a stable node.
The scale factor shows a phantom behavior.
\begin{equation}
a(t)=a_0{|t-t_0|}^{\frac{1}{-2.1195}}.
\end{equation}
5.
\begin{equation}
(0.6979, -2.0936, 0.3742, -0.0326, -0.0122, 0)
\label{b10}
\end{equation}
Eigenvalues are
\begin{equation}
( 4.7703,-0.3742, -8.2381, -0.3742, \left\{
\begin{array}{l}
-0.5613, w=-1\\
-2.0613, w=-0.5\\
-3.4678, w=0\\
-5.0613, w=0.5\\
-6.5613,w=1
\end{array}
\right .)
\end{equation}

This is a phantom saddle with
\begin{equation}
a(t)=a_0{|t-t_0|}^{\frac{1}{-0.0936}}.
\end{equation}
We now turn to the case of
$\alpha=-1$.
Similarly as before, we get the following stationary points:
\\1.
\begin{equation}
(1, -2, 0, 0, 0, 0)
\label{b12}
\end{equation}
with eigenvalues
\begin{equation}
( -3-3w, -3.4198, 0.4198, -1.5000 - 5.5844i, -1.5000 + 5.5844i)
\label{b13}
\end{equation}
This is de Sitter point which is a saddle.

2.
\begin{equation}
\left (\frac{1}{2}\!-\!\frac{1}{6}w, \frac{1}{2}w\!-\!\frac{3}{2}, \!-\!2\!-\!2w, \frac{1\!-\!w^2}{3(w-3)}, \frac{2(w^2\!-\!1)(1\!+\!w)}{3(w-3)},  \frac{(5w^3\!-\!8w^2\!+\!20w\!+\!45)}{9(w-3)}\right )
\label{b14}
\end{equation}
At least one of eigenvalues is not negative  one
\begin{equation}
2(1+w)\geqslant 0,
\end{equation}
others are
\begin{equation}
\begin{array}{l}
\\ 5/4w-1/4 \pm
\\1/4{(-7w^2+94w-119+8{(-14w^4+34w^3+25w^2-300w+1035)}^{1/2})}^{1/2},\\
\\ 5/4w-1/4 \pm
\\1/4{(-7w^2+94w-119-8{(-14w^4+34w^3+25w^2-300w+1035)}^{1/2})}^{1/2}.
\end{array}
\end{equation}
This is a matter dominated point which is a saddle with
\begin{equation}
a(t)=a_0{|t-t_0|}^{\frac{2}{w+1}}.
\end{equation}
3.
\begin{equation}
(0.7023, -2.1069, 0.4275, 0.0374, 0.0160, 0)
\label{b16}
\end{equation}
Eigenvalues are
\begin{equation}
( -1.7672 + 5.7921i,-0.4275, -3.5343, -1.7672 - 5.7921i, \left\{
\begin{array}{l}
-0.6412, w=-1\\
-2.1412, w=-0.5\\
-3.5343, w=0\\
-5.1412, w=0.5\\
-6.6412,w=1
\end{array}
\right .)
\end{equation}

For all studied values of $w$ this point is an attractive focus.
The behavior of the scale factor is of phantom type:
\begin{equation}
a(t)=a_0{|t-t_0|}^{\frac{1}{-0.1069}}.
\end{equation}

All these points are summarize in the following two tables.

\begin{tabular} {|c|c|c|c|}
\multicolumn{4}{c}{}\\
\multicolumn{4}{c}{Table 1. Stationary points for
$f_1=R+\beta R^3$ with $\beta=1$, $f_2=\alpha R\Box R$ with $\alpha=1$.}\\
\multicolumn{4}{c}{}\\
\hline
{Number} & Coordinates of stationary & Stability & Time dependence
of\\
{of point} & points & type & scale factor $a(t)$\\
\hline
{} & {} & {} & {}\\
{} & $x=1$, & {} & {}\\
{} & $y=-2$, & {} & {}\\
{} & $z=0$, & Saddle  & {}\\
1. & $m=0$, & {} & $a(t)=a_0{e}^{\frac{t-t_0}{\sqrt12}}$\\
{} & $n=0$, & {} & {}\\
{} & $\Omega=0$. & {} & {}\\
{} & {} & {} & {}\\
\hline
{} & {} & {} & {}\\
{} & $x=1.3732$, & {} & {}\\
{} & $y=-4.1195$, & {} & {}\\
{} & $z=8.4778$, & Attractive node & {}\\
2. & $m=-1.0700$, &  {} & $a(t)=a_0{|t-t_0|}^{-0.4718}$\\
{} & $n=-9.07100$, & {} & {}\\
{} & $\Omega=0$. & {} & {}\\
{} & {} & {} & {}\\
\hline
{} & {} & {} & {}\\
{} & $x=0.1623$, & {} & {}\\
{} & $y=-0.4870$, & {} & {}\\
{} & $z=-6.0520$, & Saddle & {} \\
3. & $m=-1.0625$, & {} & $a(t)=a_0{|t-t_0|}^{0.6609}$\\
{} & $n=6.4305$, & {} & {}\\
{} & $\Omega=0$. & {} & {}\\
{} & {} & {} & {}\\
\hline
{} & {} & {} & {}\\
{} & $x=0.6979$, & {} & {}\\
{} & $y=-2.0936$, & {} & {}\\
{} & $z=0.3742$, & Saddle & {}\\
4. & $m=-0.0326$, & {} & $a(t)=a_0{|t-t_0|}^{-10.6724}$\\
{} & $n=-0.0122$, & {} & {}\\
{} &$\Omega=0$. & {} & {}\\
{} & {} & {} & {}\\
\hline
{} & {} & {} & {}\\
{} & $x=\frac{1}{2}-\frac{1}{6}w$, {} & {}\\
{} & $y=-\frac{3}{2}+\frac{1}{2}w$, & {} & {}\\
{} & $z=-2-2w$, & Saddle & {}\\
5. & $m=\frac{1}{3}\frac{-1+w^2}{-3+w}$, & {} &
$a(t)=a_0{|t-t_0|}^{\frac{2}{w+1}}$\\
{}  & $n=-\frac{2}{3}\frac{(-1+w^2)(1+w)}{-3+w}$, & {} & {}\\
{} & $\Omega=-\frac{1}{9}\frac{5w^3+22w^2-34w-63}{w-3}$. & {} & {}\\
{} & {} & {} & {}\\
\hline
\end{tabular}
\\\rule {0pt}{10mm}
\\
\begin{tabular} {|c|c|c|c|}
\multicolumn{4}{c}{}\\
\multicolumn{4}{c}{Table 2. Stationary points for
$f_1=R+\beta R^3$, $\beta=$, $f_2=\alpha R\Box R$ with $\alpha=-1$.}\\
\multicolumn{4}{c}{}\\
\hline
{Number} & Coordinates of stationary & Stability & Time dependence
of\\
{of point} & points & type & scale factor $a(t)$\\
\hline
{} & {} & {} & {}\\
{} & $x=1$, & {} & {}\\
{} & $y=-2$, & {} & {}\\
{} & $z=0$, & Saddle & {}\\
1. & $m=0$, & {} & $a(t)=a_0{e}^{\frac{t-t_0}{\sqrt12}}$\\
{} & $n=0$, & {} & {}\\
{} & $\Omega=0$. & {} & {}\\
{} & {} & {} & {}\\
\hline
{} & {} & {} & {}\\
{} & $x=0.7023$, & {} & {}\\
{} & $y=-2.1069$, & {} & {}\\
{} & $z=0.4275$, & Attractive node & {}\\
2. & $m=0.0374$, & {} & $a(t)=a_0{|t-t_0|}^{-9.3545}$\\
{} & $n=0.0160$, & {} & {}\\
{} & $\Omega=0$. & {} & {}\\
{} & {} & {} & {}\\
\hline
{} & {} & {} & {}\\
{} & $x=\frac{1}{2}-\frac{1}{6}w$, {} & {}\\
{} & $y=-\frac{3}{2}+\frac{1}{2}w$, & {} & {}\\
{} & $z=-2-2w$, & Saddle & {}\\
3. & $m=-\frac{1}{3}\frac{-1+w^2}{-3+w}$, & {} &
$a(t)=a_0{|t-t_0|}^{\frac{2}{w+1}}$\\
{}  & $n=\frac{2}{3}\frac{(-1+w^2)(1+w)}{-3+w}$, & {} & {}\\
{} & $\Omega=-\frac{1}{9}\frac{5w^3-8w^2+20w+45}{w-3}$. & {} & {}\\
{} & {} & {} & {}\\
\hline
\end{tabular}
\\
\\
It is important to compare these results with those found in the absence of $f_2$- term.
For the function $f_1=R+R^3$ these are three stationary points -- de Sitter (which is
an unstable node), a stable vacuum power-law solution, a solution
with  $a \sim t^{1/2}$ behavior and an unstable matter-dominated
power-law solution (the latter case represents a situation where the power index
in not changed due to the dalambertian term).

In the presence of the dalambertian term we have analogs of three of these four solution.
It is interesting, that substituting in the general form of equations of motion (16)
the function $f_1=R^3$ and using
$f_{1R}=3R^2, f_{1RR}=6R, f_{1RRR}=6, b=2, c=1$, and, correspondingly,
$y=-3x, r=-z^2 \alpha/72$,
which gives the equation of motion in the form

\begin{equation}
\begin{array}{l}
x'=\frac{3xz}{2}+x(-6x+4-z)\\
\\
z'=-\frac{z^2}{2}-54\frac{xm}{\alpha}-z(1+3x)\\
\\
m'=n-mz\\
\\
n'=
-\frac{z^2}{2}+54\frac{xm}{\alpha}+2x+6xm-\frac{\alpha z^2}{36}+\\
\\+(-2x+z+1+m(-3x+1)+n-\frac{z^2\alpha}{72})(1-3w)-n(1+3x+z).
\end{array}
\end{equation}
we get a bit different set of stationary points. As this theory can be treated as high-curvature
limit of the $R+R^3$ theory when Einstein contribution can be neglected, we
do not find de Sitter point. However, this is not the only difference. There are two
additional solutions which have the following form for $\alpha=1$ for vector $(x,z,m,n,\Omega)$:
\begin{equation}
(0,  0,  -1, 0, 0)
\end{equation}
with eigenvalues
\begin{equation}
(4,1-3w,-1, -1)
\end{equation}
and
\begin{equation}
(0,-2, -\frac{19}{18}, \frac{19}{9}, 0)
\end{equation}
with eigenvalues
\begin{equation}
(3,3-3w,-1,-1)
\end{equation}
which correspond to $a \sim t^{1/2}$ behavior. Both solutions have $x=0, y=0$, and, so,
they can not be found between solutions of Eqs. (19) due to the factor $x+y$ in denominator.
This is the same reason why we do not find a solution corresponding to a Friedmann
Universe in low-curvature limit of the theory $R+R^3$ -- this is already remarked in .

For other values of $\alpha$ coordinates of one of these point changes, though
the behavior $a \sim t^{1/2}$ and the eigenvalues remain unchanged which indicates that
these two points are saddles.

Returning to other points,
we can see that the difference between $\alpha=1$ and $\alpha=-1$ cases is in number
of power-law vacuum solutions -- there are three of them for $\alpha=1$ (with one
stable solution) and one (which is stable) for $\alpha=-1$. We remind a reader that
for $\alpha=0$ we have one stable node $a \sim (t-t_0)^{-10}$. Using power-law
ansatz we can study a general case.

Substituting
$a(t)=a_0{|t-t_0|}^{\frac{1}{A}}$ into eq. (9),we get
for  $f_1=R+\beta R^3$
\begin{equation}
\begin{array}{l}
\frac{1}{A^6{(t-t_0)}^6}(\beta(-6)^3{(2-A)}^3+
3(A^4{(t-t_0)}^4+3\beta(-6)^2{(2-A)}^2)+\\
\\+18\beta(-6)12{(2-A)}^2A+\alpha(36(-6){(1-A)}^2(2-A)A+\\
\\+6\alpha(-144)(2-A)(1-A)A^2-72\alpha{(2-A)}^2A^2-K{\rho}_0A^6{(t-t_0)}^{-\frac{3(1+w)}{A}+6})=0.
\end{array}
\end{equation}
Neglecting matter and Einstein term we get the equation for $A$
\begin{equation}
20\alpha A^3+2A^2(15\beta-8\alpha)-A(57\beta+6\alpha)-6\beta=0.
\label{d1}
\end{equation}
The solution is plotted in Fig.1. Stability analysis shows that one solution is stable,
stable branches are shown in bold.

Other solutions do not show any significant dependence on the value of $\alpha$.
The coordinates of de Sitter point do not depend on $\alpha$, and stability analysis shows
that it is always a saddle. For the matter dominated point one eigenvalue is equal
to $2-2w$ for any $\alpha$, so this point is unstable (though dimension of unstable
manyfold for this point does depend on $\alpha$). Finally, eigenvalues of the points
with $a \sim t^{1/2}$ does not depend on $\alpha$.

\begin{figure}[h]
\begin{center}
\includegraphics[width=0.6\textwidth]{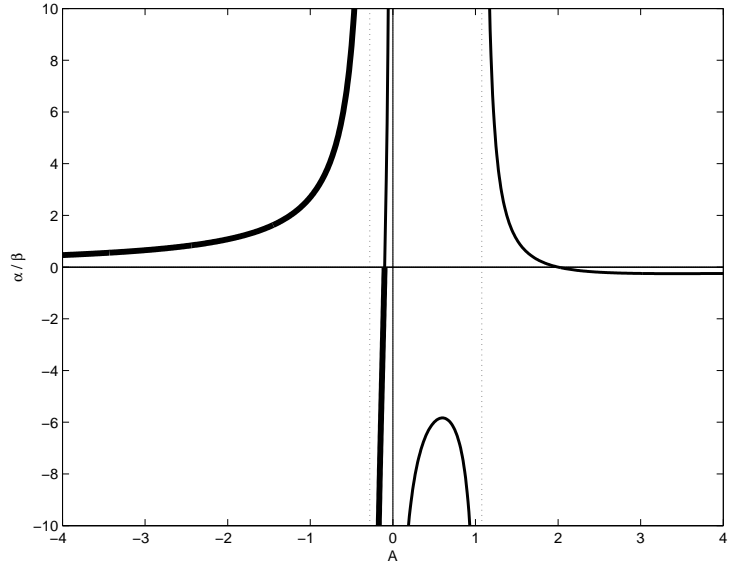}
\caption{Power index $A$ of a vacuum power-law solution depending on coupling
constants $\alpha$ and $\beta$.Stable branches are  bold lines.}
\end{center}
\end{figure}

\section {The case of $f_1=R+R^N, N \ne 3$}

The equation (49) shows also why the case of $f_1=R+R^3$ is exceptional for theories
with $f_2=R\Box R$. In this case for power-law ansatz $a \sim (t-t_0)^{1/A}$ terms
originating
from $f_1$ and $f_2$ are equally important. For  functions $f_1=R+R^N$ with $N>3$
in the high-curvature regime terms from $f_1$ dominates (in low-curvature
regime dominant terms are those from Einstein part of $f_1$). This means that
influence of $f_2$ could be important in some transient regime of intermediate energy,
and, so, it can not be found by studies of stable points and corresponding
asymptotic regimes.

On the other hand, for $1<N<3$ the evolution of the Universe begins with
$R\Box R$ dominant regime, after that $R^N$ corrections can be important in
an intermediate regime, and for low-curvature stage Einstein term becomes
dominant.  To study such type of a transient dynamics we need other methods.

In what follows we consider two particular problems which can be solved
using a specially chosen anzatz.

\subsection {Stability of de Sitter solution}

It is known that in a modified gravity with $f_1=R+R^N$ de Sitter solution exists
and stable for $0<N<2$, $N \ne 1$. We study what happens if we include the dalambertian term.

In this section it will be better for us to use the different signature from previous one. Thus we put
\begin{equation}
\mathrm{g}_{ik}=diag(-1,a^2,a^2,a^2)
 \label{1.2}
\end{equation}
In this notation some values change the sign, for instance we have $R=6(\dot H+2H^2)$, $R_{00}=3(\dot H +H^2)$ and respectively $\Box R = -\ddot R - 3H\dot R$.

We take function $f$ in the form:

\begin{equation}
f=R+\beta R^N + \alpha R\Box R.
 \label{2.1}
\end{equation}
Thus $00$ component of (\ref{a1}) takes the form:

\begin{equation}
\begin{array}{l}
6H^2+ \beta [ (1-N)R^N + 6H^2NR^{N-1} + 6HN(N-1)R^{N-2}\dot R ]\\
\\
 +\alpha[2R\ddot R +36 H^3 \dot R -\dot R^2 -48 H^2 \ddot R - 12H\dddot R]=0.
\end{array}
 \label{2.2}
\end{equation}
This representation is more comfortable for following investigations. There is a dS-solution in the equation (\ref{2.2}):
\begin{equation}
R_0^{N-1}=\frac{1}{\beta (N-2)}.
 \label{2.3}
\end{equation}
Note that here we have $R_0=12H_0^2$. We can see that there is no dS-solution for $N=1$ in empty space, as well as there is no dS-solution for $N=2$ in empty space. For $N=3$ we have dS-solution $H_0^2=1/(12\sqrt{\beta})$. Also its need to note that dS-solution exist for positive $\beta$ when $N>2$ and for negative one when $N<2$. This result is the same as for usual $f(R)$-gravity, because additional term $\alpha R\Box R$ can not produce new dS-solution since it contain only curvature derivative, however it may change the stability of dS-solutions due to terms with
higher time derivatives.

   For finding stability conditions we study small perturbations near dS-solution $H_0$. We note  that $H=H_0+\delta H$, $R=R_0+\delta R= R_0+6(\delta\dot H + 4 H_0\delta H)$ and $R^N=R_0^N+6NR_0^{N-1}(\delta\dot H + 4H_0\delta H)$, where we keeping only linear with respect to $\delta H$ terms. Substituting this relations into (\ref{2.2}) and keeping only linear terms with respect to $\delta H$ terms and using (\ref{2.3}) we find
\begin{equation}
\begin{array}{l}
24H_0\alpha\delta H^{(4)} + 12\alpha R_0\delta \dddot H + \left (10H_0R_0\alpha -\frac{N(N-1)}{N-2}H_0^{-1} \right )\delta\ddot H\\
\\
 - \left ( 2R_0^2\alpha +3\frac{N(N-1)}{N-2} \right )\delta\dot H + 4H_0(N-1)\delta H=0.
\end{array}
 \label{2.4}
\end{equation}

We search for solutions of this equation taken in the form $\delta H=e^{\lambda t}$. After substituting this representation into equation (\ref{2.4}) it transforms to usual algebraical equation of forth order with four solution $\lambda_i$ (in the general complex case). It is clear that for stability solution $H_0$ need $Re(\lambda_i)<0$ for any $i$.
This algebraical equation is rather complicate one, so we use a special method (Routh-Hurwitz theorem), which detects if any root of this equation have a positive real part. For details see, for instance \cite{ssylka1}.
Briefly it works as follows: from coefficients of equation (\ref{2.4}) we construct some determinants $T_i$ and if all them are positive then all real part of solutions $\lambda_i$ are negative. Here we present the results of calculations of $T_i$:
\begin{equation}
T_0=24H_0\alpha,
 \label{2.5}
\end{equation}
\begin{equation}
T_1=144H_0^2\alpha,
 \label{2.6}
\end{equation}
\begin{equation}
T_2=72H_0\left (12\cdot 28H_0^4\alpha^2 - \alpha\frac{N(N-1)}{N-2}\right ),
 \label{2.7}
\end{equation}
\begin{equation}
T_3=9\cdot 24 H_0 \left ( -12^2\cdot 14\cdot 16 \alpha^3 H_0^8 - 48 \alpha^2 H_0^4\frac{N-1}{N-2}(13N-16) + \alpha  \frac{N^2(N-1)^2}{(N-2)^2}\right ) ,
 \label{2.8}
\end{equation}
\begin{equation}
T_4=4H_0(N-1)T_3.
 \label{2.9}
\end{equation}
Note also that for positive $\alpha$ we need condition $T_i>0$, but for negative one we need $T_i<0$. First of all we can see that for $N<1$ $T_3$ and $T_4$ always have opposite sing and therefor dS-solution is unstable for any $N<1$ (note that we interested in only $H_0>0$ case). However,  for sufficiently small values of $H_0$ all terms with high order $H_0$ can be neglected and all $T_i$ have the same  sign in the range $1<N<2$. It mean that in this range dS-solution may be stable for sufficiently big $|\beta |$. This analytical result is confirmed by numerical investigation (see Fig.2).

\begin{figure}[h]
\begin{center}
\includegraphics[width=0.6\textwidth]{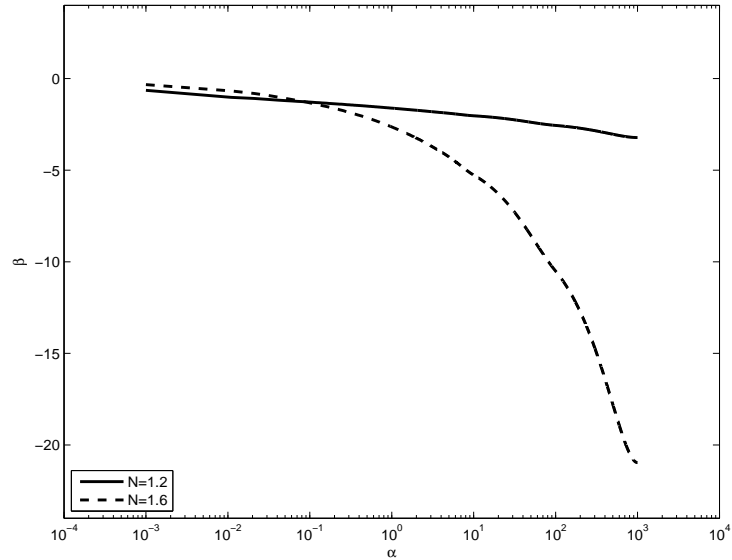}
\caption{Zones of stability of de Sitter solution for two values of $N$. The de Sitter
solution is stable below the curves.}
\end{center}
\end{figure}

\subsection {Ruzmaikin solution and its analogs}

The method we used have another limitation. Usually, having the variables
determined we obtain the time dependence of the scale factor via Eqs. (10).
The simplest way is to use equation for $y$.
If the scale factor behavior is of power-law type, this equation gives us
the exact solution. However, for other types of dynamics we can not be sure
that the set of variables can distinguish asymptotic regimes, which are
different from the cosmological point of view.  An interesting example is
the situation in modified gravity with $f_1=R+R^2$. This particular case is
an exceptional one in the sense that there are no de Sitter solution in this theory
(see above).
Instead, the Ruzmaikin solution $H \sim t$ exists \cite{Ruzmaikin}. From the
cosmological perspective, this solution describing super-inflating Universe
is very different from the de Sitter solution, however, it gives the same
values of $x, y, z$ as the latter one. Only by using initial equations (8)--(9)
we can understand that something special happens for this particular function $f_1$.
This explains why this solution have not been distinguished in papers \cite{Dunsby2, Polarski}.
It is interesting that another set of variables used in \cite{Hervik} does discriminate
between de Sitter and Ruzmaikin solutions.

In what follows we show that similar solution exists for the theory with
$f_1=R+R^2$ and $f_2=R\Box R$. Consider
\begin{equation}
\begin{array}{l}
a(t)=const\cdot e^{\frac{Bt^{n+1}}{n+1}},
\\H=Bt^n.
\end{array}
\end{equation}
we get
\begin{equation}
\dot H=Bnt^{n-1},
\end{equation}

\begin{equation}
R=-6(2B^2t^{2n}+Bnt^{n-1}),
\end{equation}

\begin{equation}
\dot R=-6(4nB^2t^{2n-1}+Bn(n-1)t^{n-2}),
\end{equation}

\begin{equation}
\ddot R=-6(4n(2n-1)B^2t^{2(n-1)}+Bn(n-1)(n-2)t^{n-3}),
\end{equation}

\begin{equation}
\Box R=
-72nB^3t^{3n-2}-6n(11n-7)B^2t^{2(n-1)}-6Bn(n-1)(n-2)t^{n-3},
\end{equation}

\begin{equation}
\frac{\partial\Box R}{\partial t}\!=\!
-72n(3n\!-\!2)B^3t^{3(n-1)}\!-\!12n(11n\!-\!7)(n\!-\!1)B^2t^{2n-3}\!-\!6Bn(n\!-\!1)(n\!-\!2)(n\!-\!3)t^{n-4}.
\end{equation}

Equations of motion in the absence of matter take the form
\begin{equation}
\begin{array}{l}
Et^{3n-1}+Mt^{2(n-1)}+St^{2n}+\\
\\+D_1t^{5n-1}+E_1t^{4n-2}+S_1t^{3(n-1)}+M_1t^{2(n-2)}=0,
\end{array}
\end{equation}
where constant coefficients depend on  coupling constants. We need two of them
\begin{equation}
\begin{array}{l}
S=3B^2,\\
\\
D_1=36\cdot12\alpha B^5n.\\
\end{array}
\end{equation}
We denote the terms originating from the dalambertian by
variables with subscripts. The biggest power indexes (which we should equate) belong to the terms
$St^{2n}$ and $D_1t^{5n-1}$.
This means that
\begin{equation}
\begin{array}{l}
2n=5n-1,\\
3n=1,\\
n=\frac{1}{3},
\end{array}
\end{equation}
and we get the solution of Ruzmaikin type:
\begin{equation}
\begin{array}{l}
a(t)=const\cdot e^{\frac{3Bt^{\frac{4}{3}}}{4}},
\\H=Bt^\frac{1}{3}.
\end{array}
\end{equation}
As the original Ruzmaikin solution, this solution corresponds to $(x,y,z,m,n,r)$:
\begin{equation}
(1,-2,0,0,0,0)
\end{equation}
and is undistinguishable from de Sitter point from the viewpoint of our set of
dimensionless variables.

As the coefficient $D_1$ comes from the dalambertian contribution, this solution exists
also in a theory with $f_1=R$, $f_2=R \Box R$, i.e. theory with only dalambertian
correction to Einstein gravity.
\\

As for power-law solutions, substituting
$a(t)=a_0{(t-t_0)}^{\frac{1}{A}}$
we can obtain solutions valid in the regime when $R \Box R$ dominates.
There are  vacuum solutions with
$A_1=-0.2782$  and
$A_2$=1.0782 (these solutions represent solutions of Eq.(50)
 in the limit $\alpha \to 0$),
and matter-dominated solution
$A=\frac{1}{2}(1+w)$.
These three solutions are valid in the high-energy
 regime for  $f_1=R+R^N$, $1<N<3$.

\section{Conclusion}
We have considered cosmological dynamics in the modified
gravity with both $f(R)$ and $R \Box R$ terms
in the action. In this theory the case of $f(R)=R+R^3$ appears
to be a special one. In this particular case terms in the equations
of motion originating from these two different modifications
of Einstein gravity (namely, adding $R^3$ and $R \Box R$ to the action)
are equally important on power-law solutions. This allows us to
obtain several exact solutions and study their stability using
dynamical systems approach. Our results show that the general
picture does not change much in comparison with the known
case of $R+R^3$ gravity  -- only one stationary point (which
corresponds to a phantom solution) is stable, others (their
number can vary) are unstable.

Other power-law functions $f(R)=R+R^N$ lead to zones of $f(R)$
dominance (near a cosmological singularity for $N>3$), or $R \Box R$
dominance ($R^N$ corrections are sub-dominant in comparison with
those from the dalambertian near a cosmological singularity if
$N<3$), as well as existence of transient epoch when some terms
can be temporarily important ($R^N$ corrections for $N<3$ and
$R \Box R$ corrections for $N>3$).

However, this does not mean that sub-dominant terms in some particular
epoch are absolutely unimportant. The reason of this is
extra derivatives in terms originating from $R \Box R$ contribution.
This means that some stable points existing in $R+R^N$ gravity
can loose their stability due to $R \Box R$ contribution.
We study in detail one particular example of this situation, namely
stability of de Sitter solution. It is known that de Sitter solution
exists and stable in $f(R)$ gravity with $f=R+R^N, N<2\, (N\neq 1)$. $R \Box R$
terms vanishes on de Sitter, however, extra dimensions of the phase
space may violate stability. We have shown that, indeed, for $N<1$
the de Sitter solution is unstable (for $1<N<2$ the stability
depends on coupling constants).

Finally, a modification of Ruzmaikin solution for $R+R^2$ gravity
valid for theories with $R \Box R$ corrections is found.

\section*{Acknowledgments}
This work was  supported by the RFBR grants 08-02-00923 and
09-02-12417-ofi-m. Authors are grateful to Alexey Starobinsky and David
Polarski for discussions.

\end{document}